# Observation of Topological Nodal-Ring Phonons in Monolayer Hexagonal Boron Nitride

Zhiyu Tao(陶志禹)[1,6#], Yani Wang(王雅妮)[2,4#], Shuyi He(何舒怡)[3#], Jiade Li(李佳德)[1], Siwei Xue(薛思玮)[5,1], Zhibin Su(苏智斌)[1,6], Jiatao Sun(孙家涛)[3*], Hailin Peng(彭海琳)[2,4*], Jiandong Guo(郭建东)[1,6*], & Xuetao Zhu(朱学涛)[1,6*]

[1] Beijing National Laboratory for Condensed Matter Physics and Institute of Physics, Chinese Academy of Sciences, Beijing 100190, China

[2] Center for Nanochemistry, Beijing Science and Engineering Center for Nanocarbons, Beijing National Laboratory for Molecular Sciences, College of Chemistry and Molecular Engineering, Peking University, 100871 Beijing, China

[3] School of Integrated Circuits and Electronics, MIIT Key Laboratory for Low-Dimensional Quantum Structure and Devices, Beijing Institute of Technology, Beijing 100081, China

[4] Beijing Graphene Institute (BGI), Beijing 100095, China

[5] Department of Physics, Fuzhou University, Fuzhou 350108, Fujian, China

[6] School of Physical Sciences, University of Chinese Academy of Sciences, Beijing 100049, China

# These authors contribute to this work.

* Corresponding authors. E-mails: jtsun@bit.edu.cn; hlpeng@pku.edu.cn; jdguo@iphy.ac.cn; xtzhu@iphy.ac.cn.

## Abstract

Topological physics has evolved from its initial focus on fermionic systems to the exploration of bosonic systems, particularly phononic excitations in crystalline materials. Two-dimensional (2D) topological phonons emerge as promising candidates for future technological applications. Currently, experimental verification of 2D topological phonons has remained exclusively limited to graphene, a constraint that hinders their applications in phononic devices. Here, we report experimental evidence of topological phonons in monolayer hexagonal boron nitride using advanced high-resolution electron energy loss spectroscopy. Our high-precision measurements explicitly demonstrate two topological nodal rings in monolayer hexagonal boron nitride, protected by mirror symmetry, expanding the paradigm of 2D topological phonons beyond graphene. This research not only deepens fundamental understanding of 2D topological phonons, but also establishes a phononic device platform based on wide-bandgap insulators, crucial for advancements in electronics and photonics applications.

**Introduction**

Topological physics has expanded its scope from fermionic to bosonic systems, notably encompassing topological phonons in crystalline materials, which have been extensively investigated recently[1-4]. To date, numerous topological phonon systems have been theoretically predicted in three-dimensional (3D) bulk materials, spanning metals[5-7], semimetals[8,9], semiconductors[10,11] and superconductors[12,13]. Moreover, catalogues of topological phonon systems, derived from computed phononic band structures, have been systematically compiled[14-16], facilitating the exploration of topological phonons within crystalline solids. Experimentally, distinctive linear crossing features observed in bulk phonon branches, identified as hallmarks of 3D topological phonons, have been detected through inelastic neutron/X-ray scattering[17-19]. Unlike conventional phonons governing lattice vibrations, topological phonons exhibit non-trivial phononic band structures, and unique protected boundary modes resilient to specific perturbations. The exploration of these topological features aim to unlock new possibilities in thermal management, energy harvesting, and phononic circuitry[4,20,21], where precise control over phonon transport and manipulation is essential. Particularly in the realm of two-dimensional (2D) materials, the studies of topological phonons have emerged as an exciting frontier[7,22-28].

In 2D systems, both theoretical predictions[23] and experimental evidence have identified topological nodal-ring and Dirac phonons in graphene[29], which remains the only experimentally confirmed 2D topological phonon system so far. Modifying the symmetry of 2D lattice can induce distinct phonon modes, leading to richer topological phenomena. The exploration of topological phonons in 2D materials with diverse physical properties and symmetries[30] is crucial for advancing understanding of low-dimensional topological phonons and their potential applications. In this context, monolayer hexagonal boron nitride (MhBN) emerges as a superior candidate to graphene. Although MhBN shares a similar hexagonal lattice structure with graphene, it is expected to exhibit distinct phonon behavior due to the breaking of inversion symmetry. Unlike graphene, which has high electrical conductivity and low dielectric properties, MhBN is insulating and possesses a high dielectric constant, making it more versatile for applications such as dielectric materials in devices and high-frequency rectifiers[31,32]. Additionally, the electronegativity difference between boron and nitrogen atoms gives MhBN microscopic polarity, which does not exist in graphene. This intrinsic polarity may enable novel optical responses, suggesting new application prospects[33,34]. Crucially, the influence of such polarity on topological phonons remains unexplored. Investigating topological phonons in MhBN could not only deepen our understanding of low-dimensional topological phenomena but also pave the way for transformative phononic devices.

In this study, we systematically investigated the topological phonon structure of MhBN using high-resolution electron energy loss spectroscopy with 2D energy-momentum mapping (2D-HREELS) [35]. Our analysis mapped the phonon dispersion across the entire Brillouin zone (BZ) and experimentally revealed two mirror ($M_z$) symmetry protected topological nodal rings. Unlike Dirac phonons in graphene, the corresponding phonon branches in MhBN exhibit an energy gap stemming from broken inversion ($P$) symmetry, giving rise to more complex physical phenomena, such as chiral phonons. Notably, despite the strong dipole scattering intensities induced by MhBN's intrinsic polarity, our high-resolution measurements clearly identify the topological features throughout the phonon dispersion. These findings not only confirm the feasibility of detecting topological phonons in polar materials but also establish a foundation for developing phonon-based devices utilizing wide-bandgap 2D semiconductors.

**Topological phonons in MhBN**

In a crystalline material, the topological properties of lattice vibrations are primarily determined by lattice symmetry. Unlike the $C_{6v}$ symmetry of graphene, the lattice structure of MhBN consists of alternating boron and nitrogen atoms, breaking the inversion ($P$) symmetry of the graphene-like hexagonal lattice. This disruption reduces the symmetry of MhBN to $C_{3v}$, while preserving the time-reversal ($T$) symmetry and the vertical mirror ($M_z$) symmetry. Symmetry analysis suggests that under the protection of $T$ and $M_z$ symmetries, phonon branch crossings in MhBN can form topological nodal rings. Each nodal ring exhibits a Berry phase difference of π inside and outside of the nodal ring. However, due to the broken $P$ symmetry, the Dirac phonons, which are protected by $P \cdot T$ symmetry in graphene, will experience degeneracy lifting and energy gap opening in MhBN. Our first-principles phonon calculations confirm this model analysis. As depicted in Fig. 1a, the longitudinal acoustic (LA) and out-of-plane optical (ZO) phonon branches intersect in the dispersion. These intersection points form a continuous crossing line, establishing nodal ring NR1 around the Γ point (Fig.2 c and e), while crossings between the transverse acoustic (TA) and ZO branches form nodal ring NR2 around the K and K' points (Fig.3 c and e).

**HREELS mapping of phonon dispersions in MhBN**

The high-quality MhBN samples (shown in Fig. S1) were grown on $Cu_{0.8}Ni_{0.2}$(111)/sapphire substrates (as depicted in Fig. 1b) using the Chemical Vapor Deposition (CVD) method (Methods). HREELS measurements were conducted to observe phonon dispersions using a system capable of 2D energy and momentum mapping[35]. The instrument achieves ultimate energy and momentum resolutions of 0.7 meV and 0.002 Å$^{-1}$, respectively[35]. All measurements were performed at room temperature with an incident beam energy of 110 eV and an incident angle of 60°. Following rigorous parameter optimization to eliminate extraneous variables, a balanced consideration of measurement accuracy and efficiency under current experimental constraints establish an instrument resolution specification of 5 meV (energy) and 0.09 Å$^{-1}$ (momentum), as quantified in Fig. S2. These resolutions represent the maximum possible uncertainties in energy and momentum scales in our measurements. These resolution thresholds prove sufficient to resolve topological phononic structures in MhBN. To ensure methodological reliability, phonon measurements were systematically conducted on 5 distinct MhBN samples under identical experimental configurations, thereby confirming data reproducibility across multiple sample batches. The measured phonon dispersions along high-symmetry directions in the BZ are shown in Fig. 1c, demonstrating excellent agreement with the calculated results in Fig. 1a. The assignments of the measured phonon modes are illustrated in Fig. S3.

The topological nature of phonons is typically indicated by crossings between phonon branches across the entire BZ. Therefore, to clearly identify the global topological features of phonons in MhBN, it is essential to map phonon dispersions throughout the entire 2D BZ, not solely along high-symmetry directions. Using a step motor to continuously adjust the in-plane azimuthal angle of the sample, as illustrated in the lower panel of Fig. 1b, we conducted 2D HREELS measurements in 5° steps, collecting 25 datasets consecutively to cover a 120° area in the BZ (some typical HREELS spectra are shown in Fig.S4). Phonon spectra in other areas of the BZ were obtained by duplicating the measured data, maintaining the $C_{3v}$ symmetry constraints. As depicted in Fig. 1d, integrating 2D HREELS phonon spectra from different angles enabled us to reconstruct the 3D phonon structure ($q_x$, $q_y$, $\hbar\omega$) of MhBN across the entire 2D BZ.

**Identification of nodal-ring phonons in MhBN**

The 3D phonon mapping shown in Fig. 1d enables us to experimentally investigate the global topological structure of NR1 and NR2. As depicted in Fig. 2a, we extract and integrated the phonon dispersions of the LA and ZO branches collected within the 120° range, and extrapolated the plots by performing $C_{3v}$ symmetry to cover the entire BZ. The experimental results clearly demonstrate that the intersections of the LA and ZO branches form a closed loop NR1 around the Γ point, marked by the bright cyan curve in Fig. 2a. A stack of energy loss curves along a specific direction, presented in Fig. 2b, unambiguously indicate the details of the crossing between the LA and ZO branches. The energy of the experimentally observed NR1 remains nearly constant across the entire BZ, concentrated around 90.5 meV. Computational results in Fig. 2c also show that NR1 is non-dispersive throughout the BZ, with an energy position of approximately 91 meV, in agreement with the experimental findings. Additionally, Fig. 2d displays the energy difference between the LA and ZO modes ($|\hbar\omega_{LA} - \hbar\omega_{ZO}|$) in the first BZ. The contour of the lightly colored lines, where the energy difference is zero, outlines the rounded hexagonal shape of NR1, matching well with corresponding computational results in Fig. 2e.

Similarly, as shown in Fig. 3a, the global nature of NR2 was experimentally captured by mapping the crossings between the TA and ZO branches. The corresponding energy loss curves and computational results are plotted in Fig. 3b and 3c, respectively. Different from the constant energy of NR1, NR2 exhibits energy dispersion ranging from 77 to 83 meV across the 2D BZ. Moreover, unlike the closed loop around the Γ point of NR1, NR2 forms loops around the K and K' points in the BZ. Since the mapping using current HREELS technique can only be obtained centered around the Γ point, by rotating the sample azimuthal angle (as shown in Fig. 1b), HREELS mapping around the BZ boundary at K or K' points here are obtained by reconstructing the measured data beyond the first BZ, following the $C_{3v}$ symmetry constraints. The path shapes of NR2, based on the energy difference between the TA and ZO branches ($|\hbar\omega_{TA} - \hbar\omega_{ZO}|$), as illustrated in Fig. 3d and 3e, effectively depict the closed-loop shape of NR2 around K and K' points. Experimental and calculated results demonstrate excellent agreement.

**The intensity analysis of the nodal-ring phonon branches**

While the global dispersion crossings of the phonon branches undoubtedly reflect the topological feature of the nodal rings, it is still essential to further check the Berry phase and the intensities of the involved phonon branches. An important step involves examining the Berry phase differences of phonon branches inside and outside of the nodal rings. As depicted in Fig. S5, we employed the linear integral method to compute the Berry phase across NR1 and NR2, revealing a phase change from 0 to π, which clearly confirms the topological nature of nodal-ring phonons. Furthermore, prior studies suggest that in the topological phonon structures formed by phonon branch crossings, the intensity of each branch should remain independent post-crossing[19]. In this study, we analyzed the scattering intensities of the LA and ZO branches around the crossing points from various angles using our experimental setup, as illustrated in Figs. 4a-e. The LA and ZO phonon branches composing NR1 maintain consistent relative intensities across the intersection (representing the inner and outer parts of the nodal ring). Specifically, the intensity of LA phonons is consistently higher (e.g., Fig. 4b) or consistently lower (e.g., Fig. 4d) than that of ZO phonons across the entire momentum range. There is no indication of intensity transfer between the LA and ZO branches after their intersection. We further corroborated these findings using the scattering theory[19] applied in HREELS to theoretically simulate the intensities of the LA and ZO branches. Aligning both energy and momentum resolutions with those used experimentally, our simulations also demonstrated an absence of spectral weight transfer post-intersection, confirming excellent agreement with experimental observations. Similarly, an analysis of peak intensities for NR2, detailed in the Fig. S6, exhibited results analogous to those of NR1.

**Discussion**

The topological phonon properties of MhBN exhibit significant distinctions from those of graphene, particularly in their symmetry-dependent characteristics. Unlike graphene, where both nodal rings are around the Γ point, MhBN features a nodal ring around the K point that extends beyond the BZ boundary. This unique nodal architecture may provide additional large-momentum electron scattering channels, potentially enhancing electron-phonon coupling. Experimentally, this imposes stringent requirements: HREELS measurements must encompass the second BZ, demanding exceptional instrumental stability and spatial resolution, which is only possible in a 2D-HREELS setup.

Moreover, the inherent inversion symmetry breaking in MhBN induces several critical emergent properties that do not exist in graphene. First, the intrinsic microscopic polarity enhances the macroscopic electromechanical responses. Studies show that coupling polar materials with electric fields amplifies polar-axis phonon scattering (increasing thermal resistance) while enabling anisotropic phonon transmission[36,37]. This supports gradient polarity engineering for directional thermal conduction, providing a foundation for thermal diodes[38,39]. Moreover, heterojunctions between *h*-BN and non-polar materials create interfacial polarity gradients that control phonon dynamics, enabling efficient interfacial thermal rectification in thermal management systems[40-42].

Second, the Dirac points in graphene[23,29] are replaced by band gaps in MhBN. Comparative experimental data between graphene and MhBN, as illustrated in Fig. S7, reveal that Dirac points DP1 and DP2 in graphene are absent in MhBN, where instead, two band gaps, Δ1 and Δ2, emerge in the phonon spectra around the K points. The nodal rings are protected by the $M_z$ symmetry in both materials. It is plausible that disrupting the $M_z$ symmetry in MhBN, possibly by employing incompatible substrates such as Ni (111), could induce out-of-plane distortions[43]. This disruption might open gaps in the nodal rings (Fig. S8), a phenomenon preliminarily observed in previous studies using conventional HREELS[44], where a slight gap was detected between the LA and the ZO phonon branches in MhBN on a Ni (111) substrate.

Finally, the opening of the energy gaps Δ1 and Δ2 in MhBN gives rise to several intriguing physical phenomena. Notably, the lifting of the degeneracy between the LO and LA vibrational modes at the K point prevents their respective vibrations from canceling each other. This results in the emergence of intrinsic chiral phonon modes[45,46], a feature absent in two-dimensional planar graphene. These chiral phonon modes exhibit anisotropic vibrational patterns and carry additional phase information. Our computational results successfully capture these distinctive vibrational characteristics in MhBN, as shown in Fig. S9. The calculated anisotropic chiral phonons are strongly correlated with the nonzero Berry curvature in MhBN, which could give rise to nontrivial transport phenomena, such as the topological phonon Hall effect[45] and anisotropic heat transport within the material[47,48].

**Outlook**

Recent studies have successfully achieved chiral phonon detection by constructing light with different handedness of polarization, thereby exciting specific chirality in phonons and enabling detection at specific energy frequencies[49,50]. However, due to the small momentum carried by photons, these optical methods can only probe phonons corresponding to channels with zero momentum transfer, and thus cannot provide information on the phonon dispersion. Meanwhile, current inelastic scattering methods, such as HREELS, are limited to determining the energy and momentum of phonon modes and cannot distinguish between the opposite directions of vibration characteristic of the chiral phonons. A significant avenue for future research involves developing phase-controllable, coherently polarized probe sources in various inelastic scattering techniques. Recent advancements include a proposal for a novel filtering device designed to generate "pinwheel" states in electrons, potentially allowing for the selective detection of specific chiral phonon modes[51]. In short, the detection of topological phonons is still in its infancy, calling for the development of more innovative technologies and methods.

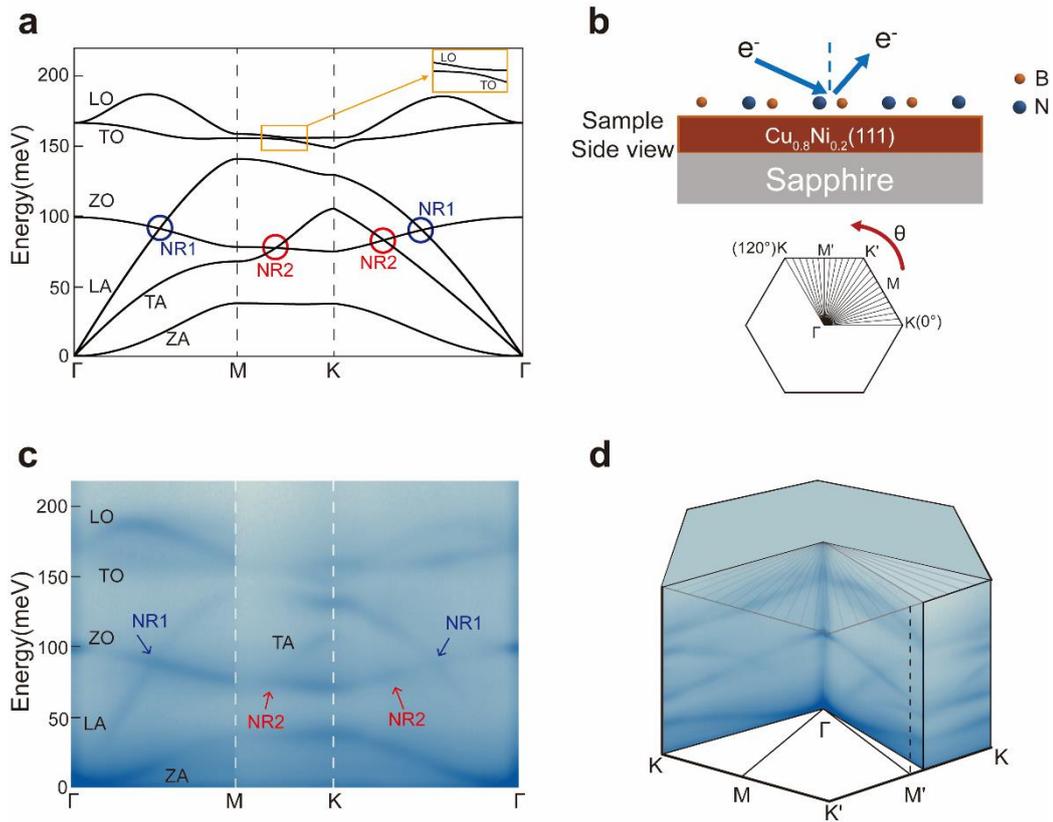

**Fig. 1.** Sample structure and phonon spectra of MhBN. (a) Calculated MhBN phonon spectra along the high-symmetry paths. The inset in the yellow box shows a local magnification of the LO and TO modes on the M-K path, where the two modes do not cross. (b) top: Side view of the MhBN sample structure, bottom: BZ of MhBN, with black radial lines indicating the scattering angles measured in the HREELS experiment. (c) Experimental MhBN phonon spectra corresponding to (a). The topological nodal ring phonons are indicated by the circles(a) and arrows(c), respectively. (d) 3D HREELS mapping of the MhBN phonon spectra.

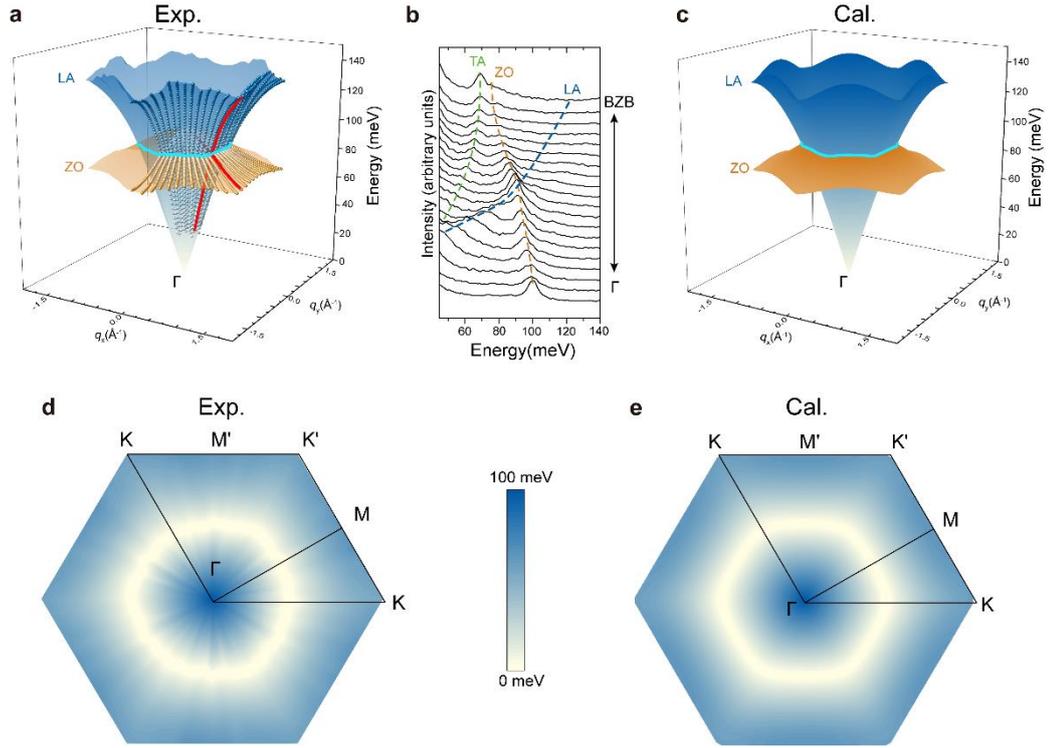

**Fig. 2.** Nodal ring phonons formed by LA and ZO branches. (a) HREELS measurements of NR1 from the dispersion of LA and ZO branches in the 2D BZ. The experimental data are marked by the color dots and the 3D colored dispersion surface is obtained by linear interpolation between experimental data points. (b) The energy distribution curves (EDC) stack of the HREELS data corresponding to the angle indicated by the red curves in (a). (c) The calculation results corresponding to (a). (d) Projected shape of NR1 obtained from the measured energy gap between the LA and ZO branches in the first BZ. (e) The calculation results corresponding to (d).

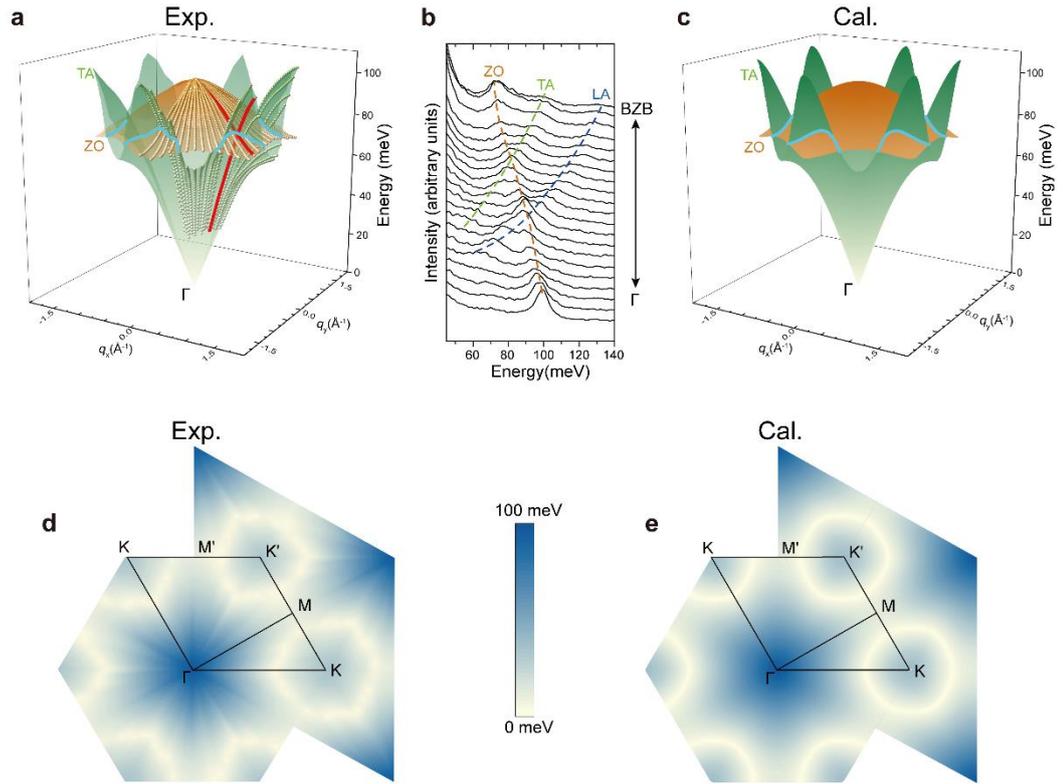

**Fig. 3.** Nodal ring phonons formed by TA and ZO branches. (a) HREELS measurements of NR2 from the dispersion of TA and ZO branches in the 2D BZ. The experimental data are marked by the color dots and the 3D colored dispersion surface is obtained by linear interpolation between experimental data points. (b) The energy EDC stack of the HREELS data corresponding to the angle indicated by the red curves in (a). (c) The calculation results corresponding to (a). (d) Projected shape of NR2 obtained from the measured energy gap between the TA and ZO branches in the first BZ. (e) The calculation results corresponding to (d).

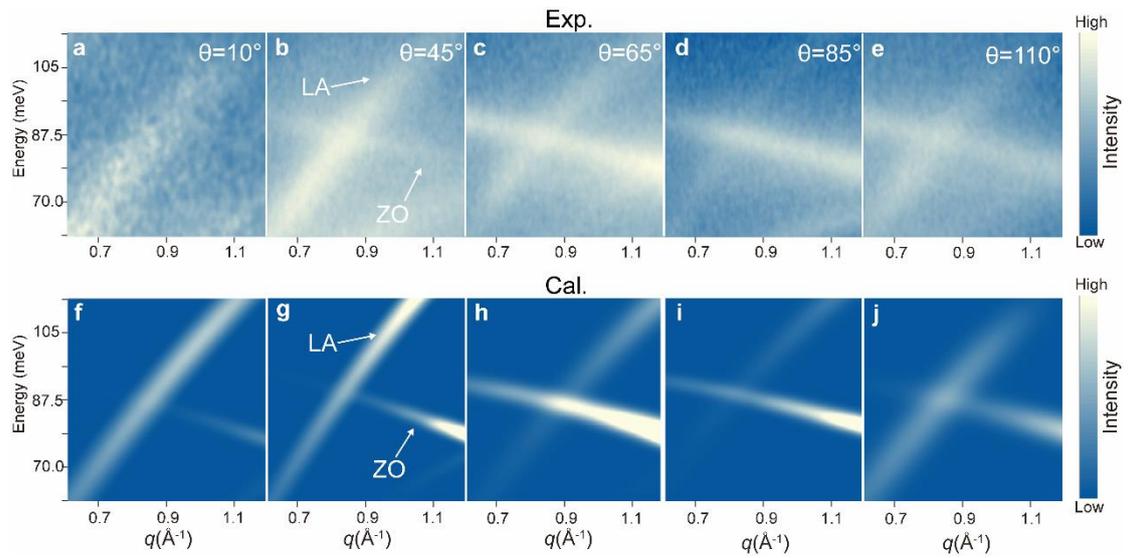

**Fig. 4.** The intensity variation of LA and ZO branches near NR1. (a-e) Intensity plots of LA and ZO branches in NR1 from 2D HREELS experiments at various azimuthal angles of the sample. The LA and ZO branches vary in intensity at different azimuthal angles. (f-j) The calculation results of impact scattering intensities for HREELS spectra corresponding to azimuthal angles in (a-e), respectively.

**Acknowledgements**

This work was supported by the National Key R&D Program of China (Grants No. 2022YFA1403000, X.Z.; No. 2021YFA1400200, X.Z.; No. 2022YFA1204900, H.P.; and No. 2020YFA0308800, J.S.), the National Natural Science Foundation of China (Grants No. 12274446, X.Z.; No. 12374172, J.S.; No. 11974045, J.S.; No. 61888102, J.S.), and the Strategic Priority Research Program of the Chinese Academy of Sciences (Grant No. XDB33000000, J.G. & X.Z.).